\documentclass{emulateapj}

\usepackage{graphicx}
\usepackage{epstopdf}

\slugcomment{Received 2009 November 14; accepted 2009 December 16 }

\shorttitle{Long Cadence \emph{Kepler} Data}
\shortauthors{Jenkins et al.}

\begin{document}


\title{Initial Characteristics of \emph{Kepler} Long Cadence Data \\
    For Detecting Transiting Planets}


\author{Jon M. Jenkins$^1$}
\author{Douglas A. Caldwell$^1$}
\author{Hema Chandrasekaran$^1$}
\author{Joseph D. Twicken$^1$}
\author{Stephen T. Bryson$^2$}
\author{Elisa V. Quintana$^1$}
\author{Bruce D. Clarke$^1$}
\author{Jie Li$^1$}
\author{Christopher Allen$^3$}
\author{Peter Tenenbaum$^1$}
\author{Hayley Wu$^1$}
\author{Todd C. Klaus$^1$}
\author{Jeffrey Van Cleve$^1$}
\author{Jessie A. Dotson$^2$}
\author{Michael R. Haas$^2$}
\author{Ronald L. Gilliland$^4$}
\author{David G. Koch$^2$}


\author{William J. Borucki$^2$}

\email{Jon.Jenkins@nasa.gov}
\affil{$^1$SETI Institute/NASA Ames Research Center, M/S 244-30, Moffett Field, CA 94035, USA}
\affil{$^2$NASA Ames Research Center, M/S 244-30, Moffett Field, CA 94035, USA}
\affil{$^3$Orbital Sciences Corporation/NASA Ames Research Center, M/S 244-30, Moffett Field, CA 94035, USA}
\affil{$^4$Space Telescope Science Institute, Baltimore, MD 21218, USA}




\begin{abstract}
The \emph{Kepler Mission} seeks to detect Earth-size planets transiting solar-like stars in its $\sim$115 deg$^2$ field of view over the course of its 3.5 year primary mission by monitoring the brightness of each of $\sim$156,000 Long Cadence stellar targets with a time resolution of 29.4 minutes. We discuss the photometric precision achieved on timescales relevant to transit detection for data obtained in the 33.5-day long Quarter 1 (Q1) observations that ended 2009 June 15. The lower envelope of the photometric precision obtained at various timescales is consistent with expected random noise sources, indicating that \emph{Kepler} has the capability to fulfill its mission. The \emph{Kepler} light curves exhibit high precision over a large dynamic range, which will surely permit their use for a large variety of investigations in addition to finding and characterizing planets. We discuss the temporal characteristics of both the raw flux time series and the systematic error-corrected flux time series produced by the \emph{Kepler} Science Pipeline, and give examples illustrating \emph{Kepler's} large dynamic range and the variety of light curves obtained from the Q1 observations.
\end{abstract}


\keywords{techniques: photometric --- methods: data analysis}



\section{Introduction}
The \emph{Kepler Mission} is designed to be capable of finding $\sim$50 Earth-like planets transiting solar-like stars in the solar neighborhood if such planets are common \citep{borucki2010}. Specifically,  \emph{Kepler} is required to obtain a signal-to-noise ratio (S/N) of 4 $\sigma$ for an 84 ppm deep, 6.5-hr transit of a G2V star with a \emph{Kepler} magnitude ($Kp$) of 12  \citep{koch2010}. This implies that the rms noise on 6.5-hr intervals must be below 20 ppm, including contributions from stellar variability. Since 2009 May 12, \emph{Kepler} has been observing $\sim$156,000 stars at 29.4 minute intervals as Long Cadence (LC) targets for the primary purpose of detecting transiting planets. Of these targets, 512 are also sampled at 1 minute intervals to support asteroseismic characterization of planetary host stars \citep{gilliland2010a, gilliland2010b}, as well as precise timing of transits. 

In Sectoin \ref{s:CDPP} we examine the photometric precision obtained to date on the first 33.5 days of data obtained during Quarter 1 (Q1) which began May 13 and ended June 15\footnote{\emph{Kepler's} science operations are broken into quarterly segments, each of which is  $\sim$93 days long and is subdivided into three monthly segments. Data are downlinked at these monthly intervals \citep{haas2010}.}. We discuss temporal characteristics of the raw and systematic error-corrected flux time series  in Section \ref{s:temporalCharacteristics}, and present sample light curves in Section \ref{s:lightCurves}. Summary remarks appear in Section \ref{s:conclusions}.

\section{Photometric Precision}\label{s:CDPP}
To address the question of the photometric precision obtained from the \emph{Kepler} LC data, we examined systematic error-corrected flux time series furnished by the Pre-search Data Conditioning (PDC) component of the Science Operations Center (SOC) pipeline\footnote{Details of the SOC science pipeline are discussed in \citet{jenkins2010}.}, hereafter called PDC time series.  At the end of each quarter, the spacecraft rolls $90^{\circ}$ about the boresight in order to reorient the solar arrays toward the Sun, and the target stars rotate onto new detectors in the Focal Plane Array \citep[FPA;][]{haas2010}. Each of these quarterly segments is processed by first calibrating the raw pixels and then the Photometric Analysis (PA) component subtracts the sky background and extracts simple aperture photometric brightness estimates for each star. Systematic errors and instrumental effects are identified and removed from the raw flux time series by PDC.

\begin{figure*}
\plotone{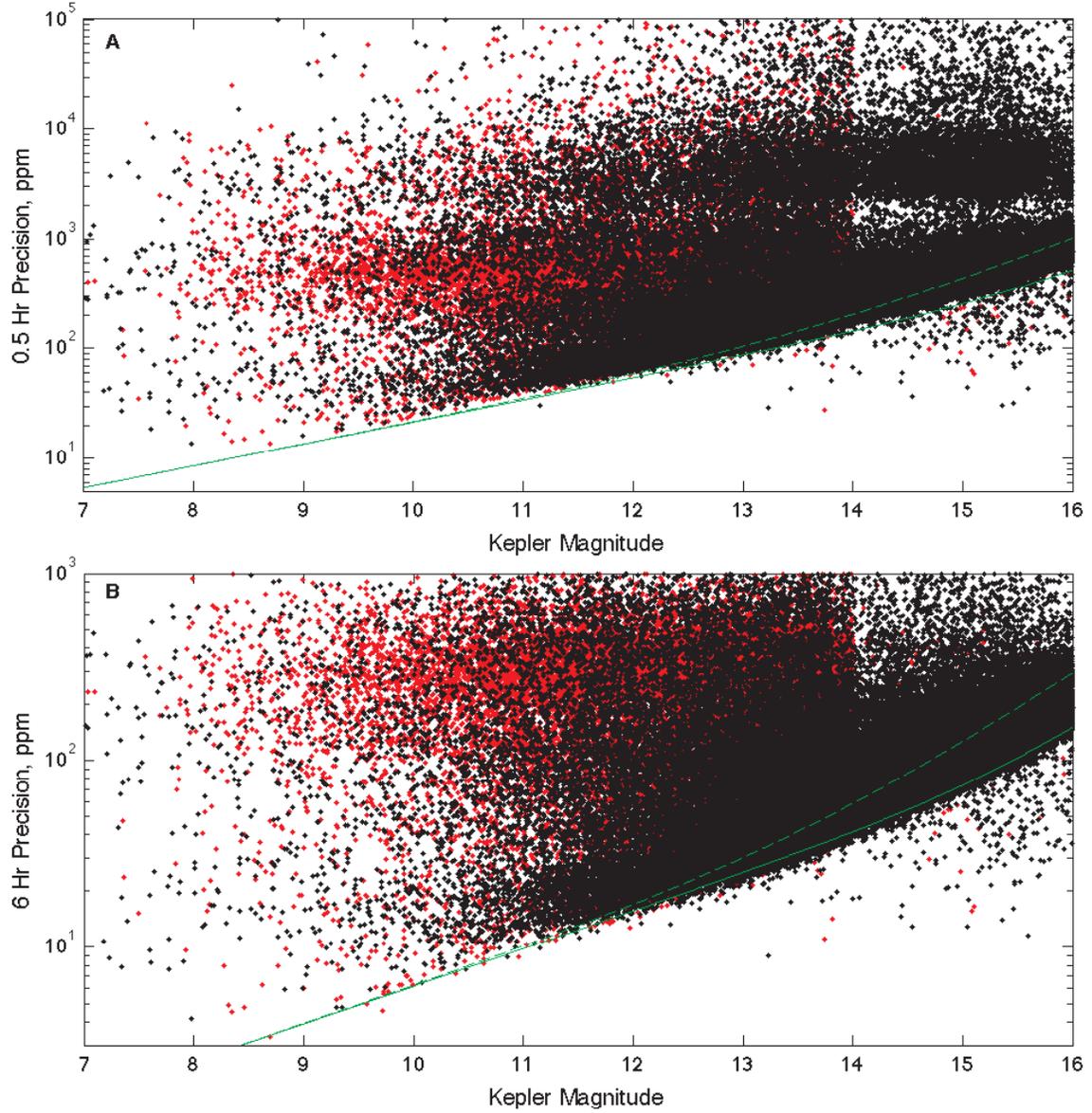}
\caption{Scatter plot of the 0.5 hr-to-0.5 hr precision (A) and the 6 hr-to-6 hr precision (B) obtained in the Q1 LC data set. The dashed and solid green curves are the upper and lower envelopes of the measurement uncertainties, respectively, propagated through the data processing steps used to construct the flux time series.  Dwarf stars (log $g \ge 4$) appear as black points, while giants (log $g < 4$) appear as red points. There is a strong separation between the dwarfs and the giants in their photometric behavior at both timescales. \emph{Kepler} is delivering near intrinsic measurement-limited noise performance across the full dynamic range of its target stars.
\label{fig:0p5hr6hrPrecision}}
\end{figure*}

The PDC flux time series for all 156,097 stars were prepared for analysis by subtracting the median flux value,  normalizing by the same, and then removing a fitted 4th order polynomial across all 1639 LC samples.  The collection of standard deviations of the normalized, detrended flux time series represents the  photometric precision on 29.4 minute intervals as a function of stellar magnitude, and this is plotted in panel A of Figure \ref{fig:0p5hr6hrPrecision}.


The standard propagation of errors is applied to each data processing step so that flux uncertainties are available along with the flux measurements themselves. These uncertainties include shot noise from photon counting statistics from the direct stellar flux, sky background and shutterless readout smear, terms such as read noise and quantization noise, and the effects of processing the data, for example, pixel-level calibrations and sky background removal. The uncertainties depend on the details of the size of the photometric aperture, the Pixel Response Function, and other considerations.  It is important to note that these uncertainties do \emph{not} include any contributions from intrinsic stellar variability. 

Plotting the individual uncertainty estimates would clutter the figure, so we fit a simple analytic expression "by eye" to the lower envelope of these uncertainties at the LC timescale: $\hat\sigma_{\mathrm{lower}} = \sqrt{c+7 \times 10^6 \max(1, Kp/14)^4}/c$ where $c = 3.46 \times 10^{[0.4*(12-Kp)+8]}$ is the number of detected $e^- $ per LC sample, and $Kp$ is the Kepler Input Catalog (KIC) magnitude for each star. The component $c$ is a Poisson term while the $(Kp/14)^4$ accounts heuristically for the increased importance of sky background and other noise sources for the dimmer stars with $Kp\ge14$. We also fit "by eye" an analytic expression to the upper envelope of this distribution as a function of $Kp$: $\hat\sigma_{\mathrm{upper}} =\sqrt{c+7 \times 10^7 }/c$. In this case a simple constant background term was sufficient. The lower and upper uncertainty envelopes appear  in Figure \ref{fig:0p5hr6hrPrecision} as solid and dashed green curves, respectively.


Three populations of stars can be seen in panel A. We take stars with KIC entries of log$g < 4$ to be giants or sub-giants\footnote{The KIC is 90\% reliable for the purposes of separating giants from dwarf stars for $Kp < 13$ and the corresponding uncertainties in log$g$ are at the 0.5 dex level  \citep{batalha2010}.}. For the most part, the main sequence stars float just above and follow the profile of  $\hat\sigma_{\mathrm{lower}}$ as a dense "finger" of points extending over the full magnitude range of the stars. The fractional width of this "finger" is approximately 1.5 near $Kp=12$ and flares out to a factor of 2 at $Kp=16$.  This is not unexpected as the dim stars are more sensitive to background sky noise, point spread function variations,  and detector properties, which vary significantly from detector to detector. A high "cloud" of stars, composed of both giants and dwarfs, hovers from 3,000 to 10,000 ppm across nearly the full magnitude range. These are variable stars of all flavors, and possibly some new ones observed for the first time. There is a sharp change in the density of points for $Kp \ge 14$  \citep{batalha2010}. This includes the fact that no stars classified as giants with $Kp>14$ were selected. Giant stars cluster on a horizontal band centered between $\sim$400 and $\sim$700 ppm with $Kp<14$. The fact that giant stars exhibit significantly more photometric variability on timescales of minutes to hours compared to dwarf stars has been previously observed by \citet{gilliland2008} in \emph{Hubble Space Telescope} observations of bulge stars, and more recently, by CoRot \citep{aigrain2009}. Giant stars universally appear to exhibit solar-like oscillations, and their increased size relative to solar-type stars causes these oscillations to occur at much lower frequencies and larger amplitudes than observed for solar-type stars \citep{gilliland2010b}. A total of 350 points lie below the  $\hat\sigma_{\mathrm{lower}}$. These stars suffer from excess flux in their apertures likely from nearby brighter stars or bleeding charge from saturated stars in the same column. When these points are plotted at their instrumental magnitudes (i.e., their measured counts are converted to magnitudes via the inverse of the relation given above), rather than by catalog magnitude, $Kp$, they are all above  $\hat\sigma_{\mathrm{lower}}$. 



As the central transit of an Earth analog lasts $\sim$13 hr, we examine the photometric precision relevant for detecting Earth-like transits lasting 6 hr or more. Each PDC flux time series considered was first high pass filtered to remove variations on timescales of 2 days and longer and then convolved with a 6-hr wide "boxcar" averaging filter. The high pass filtering was accomplished by fitting a 2nd order polynomial to a sliding analysis window 2 days wide about each point and removing the resulting polynomial at the central point for each window location. We calculated the standard deviations of the time series resulting from this process as estimates of the 6 hr precision of these data and present these in panel B of Figure \ref{fig:0p5hr6hrPrecision}. The LC uncertainty envelopes have been normalized by $\sqrt{12}$ to account for square root statistics. The giant stars (red points) occupy a band between 200 and 500 ppm. The main sequence stars continue to lie in a dense "finger" just above $\hat\sigma_{\mathrm{lower}}/\sqrt{12}$. The bottom envelope of the main sequence population tracks the expected measurement uncertainties quite well. A total of $\sim$2000 points lie below the $\hat\sigma_{\mathrm{lower}}/\sqrt{12}$, but since there are fewer independent measurements at the 6 hr level, this is expected. 90\% of the 2,665 stars with $11.5 \le Kp\le 12.5$ in the dense main sequence "finger" are below 30 ppm, 65\% are below 25 ppm, 50\% are below 23 ppm, and 31\% are below 21 ppm (which corresponds to 20 ppm at 6.5 hours). 

\emph{Kepler} is achieving extremely high precision over a remarkably large dynamic range considering that the mission design was driven largely by the 20 ppm, 6.5-hr precision requirement for $Kp$ = 12 G2V stars. Clearly, stars do exhibit stellar variability, but there are stars that fall near the $\hat\sigma_{\mathrm{lower}}/\sqrt{12}$ down to at least $Kp=9$ and there is a $Kp=8$ star exhibiting noise on 6 hr intervals of 4 ppm! We believe that further improvements in photometric precision will be made as we gain experience with the spacecraft and the photometric data.

\section{Temporal Characteristics of LC Light Curves}\label{s:temporalCharacteristics}
Although the photometric precision is a key consideration in assessing \emph{Kepler's} ability to detect transiting terrestrial planets, it is not the only one. The precision metrics discussed in Section \ref{s:CDPP} only speak to the mean noise power per 29.4 minuts or per 6 hr interval. Here we discuss temporal characteristics of the raw and PDC light curves of importance for transit detection and for non-exoplanet investigations. 

There are systematic effects that occur on timescales much longer than those relevant for detecting transits. These include Differential Velocity Aberration (DVA), which causes stars to move as much as 0.6 pixels over a 93 day period due to the large size of the field of view (FOV). \emph{Kepler} has also experienced secular pointing drifts of $\sim$ 0.05 pixels over the Q1 observations, and there have been thermal systematic effects occurring on timescales of several days and longer seasonal variations \citep{jenkins2010} as a consequence of the Sun's sky position rotating about the telescope by about $1^\circ$ day$^{-1}$. A bulk change in focus (including tip-tilt terms as well as a "piston" term) occurred in part due to desorption of water from the telescope structure and seasonal thermal effects. The gradual change in focus has caused the measured flux of stars on the outer periphery of the FOV to drop by about 0.1\% over Q1 while stars toward the interior (on the opposite side of best focus) have experienced corresponding increases in flux.

Systematic errors on these timescales are apparent especially in the raw photometric data, but are largely suppressed in the PDC time series. Astrophysical signatures occurring on timescales similar to the long term seasonal variations in \emph{Kepler's} systematic errors will inevitably be attenuated by PDC. We do not intentionally remove astrophysical signatures on any time scale. However, the conditioning of the data simply does not allow us to distinguish between astrophysical signatures and systematic errors if the former are correlated with ancillary engineering data and science data diagnostics used in the systematic error removal process. Nevertheless, the PDC time series display a remarkable wealth of astrophysics at virtually all timescales probed by the observations.

\subsection{Pointing Drift and Focus Changes}\label{ss:pointingDriftAndFocus}
\begin{figure}
\plotone{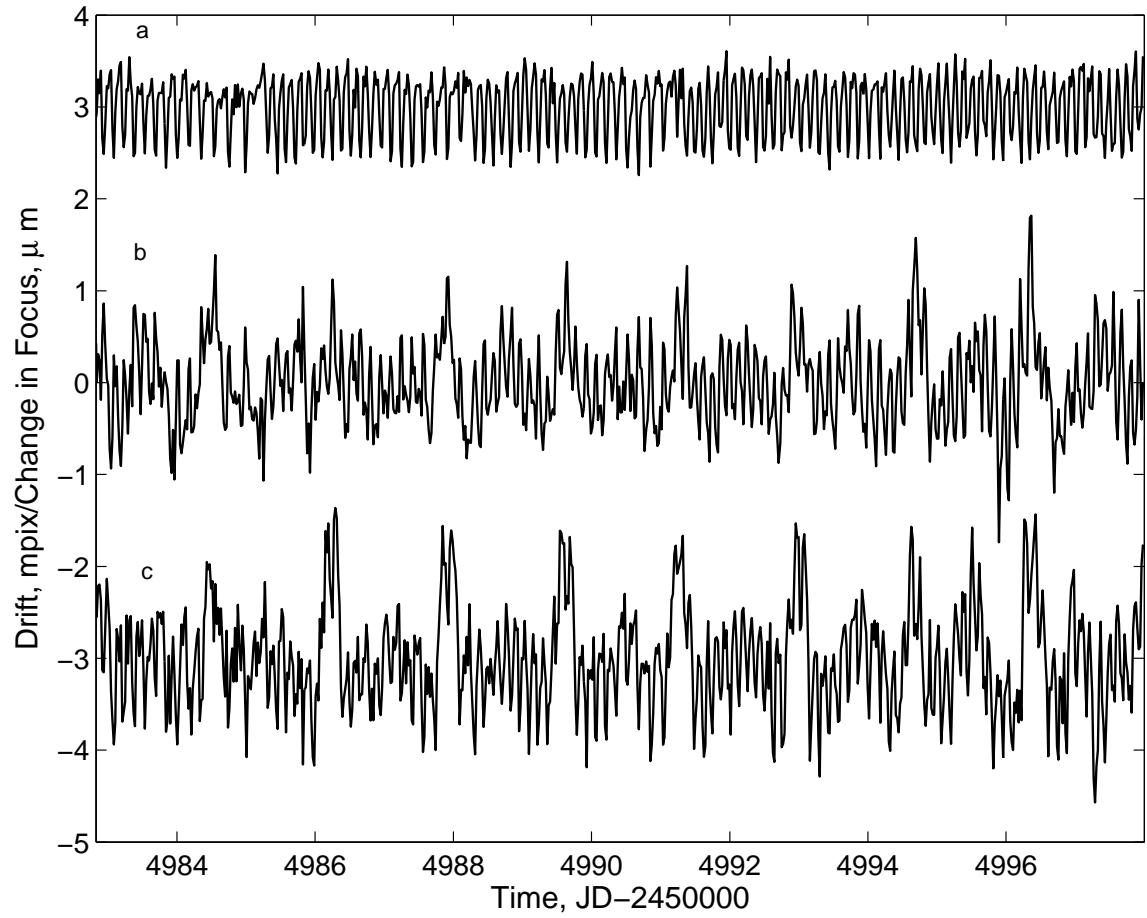}
\caption{Time series of changes in a) focus distance, b) mean row position and c) mean column position for the stars located on module 2.1 near the outside corner of the FPA over a 10 day interval during Q1 observations. 
\label{fig:platescaleRowCol}}
\end{figure}

As reported in \citet{jenkins2010} the photometer is quite sensitive even to the slightest change  in the thermal environment. Figure \ref{fig:platescaleRowCol} shows changes in a) the focus distance local to module 2.1, b) mean row, and c) mean column positions of the ensemble of stars located on module 2.1 near the outside corner of the FPA over a 15 day interval during Q1. These time series have been filtered to remove trends on timescales longer than 2 days, which are not relevant for transit detection. The focus distance of this module appears to be changing by about 1 $\mu$m every time a heater on one reaction wheel assembly turns on to keep the wheels above the setpoint temperature. These are the downward spikes in the focus distance, which represents the change in the relative distances between the stars over the observations. The mean row and column positions of these stars also experience changes in response to this heater cycle. Additionally, they experienced excursions every 1.7 days lasting about 8 hr due to an eclipsing binary on the Fine Guidance Sensor (FGS) target list that was modulating the spacecraft pointing by $\sim$1 mpix during eclipses \citep{haas2010}. 

\begin{figure}
\plotone{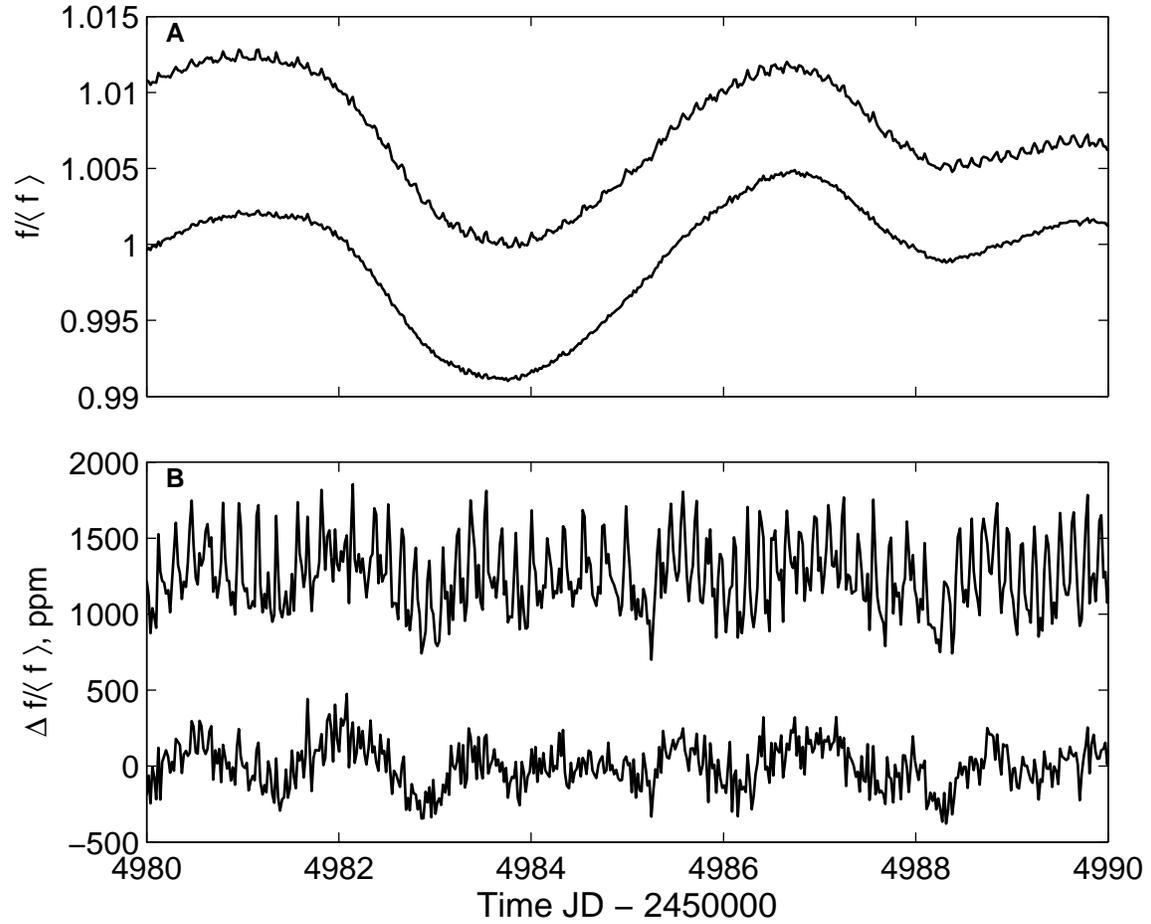}
\caption{Raw and systematic error-corrected flux time series for one star on module 2.1 observed during Q1. The curves in panel A are the relative intensity of the raw flux  (upper) and the systematic error-corrected flux  time series (lower) for this star, which exhibits variability on timescales upwards of several days. Panel B displays the results of high pass filtering each of these time series to remove long term trends. The heater cycle pulses so prevalent in the raw light curve are suppressed in the systematic error-corrected light curve, and the long term variations are preserved since they are not correlated with instrumental systematic variables.
\label{fig:paPdcSampleFlux}}
\end{figure}

Similar features appear in the raw flux time series of many LC targets. Some respond to the pointing excursions, while others do not. Most respond to focus changes at some level. Fortunately, PDC  was designed to remove systematic effects of this kind and does a good job of removing such signatures for relatively quiet stars and for stars that exhibit coherent oscillations with slowly varying frequency components. The harmonic components of the latter type star are removed prior to removing signatures correlated with information such as presented in Figure \ref{fig:platescaleRowCol}, and are then restored at the end of the correction process. There are incoherently variable targets for which it is difficult to separate the intrinsic stellar variability from systematic effects. PDC detects when this is the case and provides a much simpler detrending scheme for such targets. Figure \ref{fig:paPdcSampleFlux} exhibits one example of raw and PDC flux time series for one star located on module 2.1 which is one of the modules most sensitive to the short-term thermal effects.

We are making excellent progress with respect to understanding the behavior of the photometer. However, it is certain that much future work will consist of monitoring the systematics available through ancillary engineering data and science data diagnostics so that we can best remove the effects from the flux time series while retaining as much intrinsic stellar variability as possible.



\subsection{Cosmic Rays}\label{ss:cosmicrays}
Pre-launch expectations for the cosmic rays \emph{Kepler} would encounter on orbit were informed by a radiation model developed by Ball Aerospace Technologies Corporation (Neal Nickles, private communication 2003). Based on this model, we adopted a conservative hit rate of 5  cm$^{-2}$ sec$^{-1}$, and the mode of the distribution of total charge deposited per hit was expected to be 2500 e$^-$. These facts translate into an expectation that every pixel will receive a direct cosmic-ray hit rate of $\sim$ 3 day$^{-1}$ and that each such event would contribute charge to $\sim$4 adjacent pixels. Analysis of dark frames acquired prior to releasing \emph{Kepler's} dust cover indicate a cosmic ray deposition rate of 10 cm$^{-2}$ s$^{-1}$  \citep{vancleveandcaldwell2009}, consistent with our pre-launch expectations. Thus, cosmic rays do not pose a significant noise problem for transit detection. 

\subsection{Image Artifacts}
Image artifacts introduced by some circuits in the analog electronics for \emph{Kepler's} Focal Plane Array were identified during pre-flight testing \citep{caldwell2010}. Although a small fraction of the FOV is impacted at any one time, the effects are either amenable to correction in data processing or can be identified and excluded from consideration at little risk to the primary mission goals. The risk to the hardware and cost of attempting a fix to the electronics were too high to proceed with a direct intervention with the flight hardware. The \emph{Kepler} Science Office and SOC are engaged in developing, testing and implementing software mitigations for these image artifacts \citep{jenkins2010}. Typical amplitudes of these artifacts are $< 1$ ADU pixel$^{-1}$ read$^{-1}$, so that in most cases their presence in the flux time series of stars can be detected only by sophisticated analyses. Once the mitigations are in place we expect to recover the required photometric performance for over 90\% of the FOV \citep{caldwell2010}. 

\subsection{CCD Defects}\label{ss:ccdDefects}
\begin{figure*}
\epsscale{.75}
\plotone{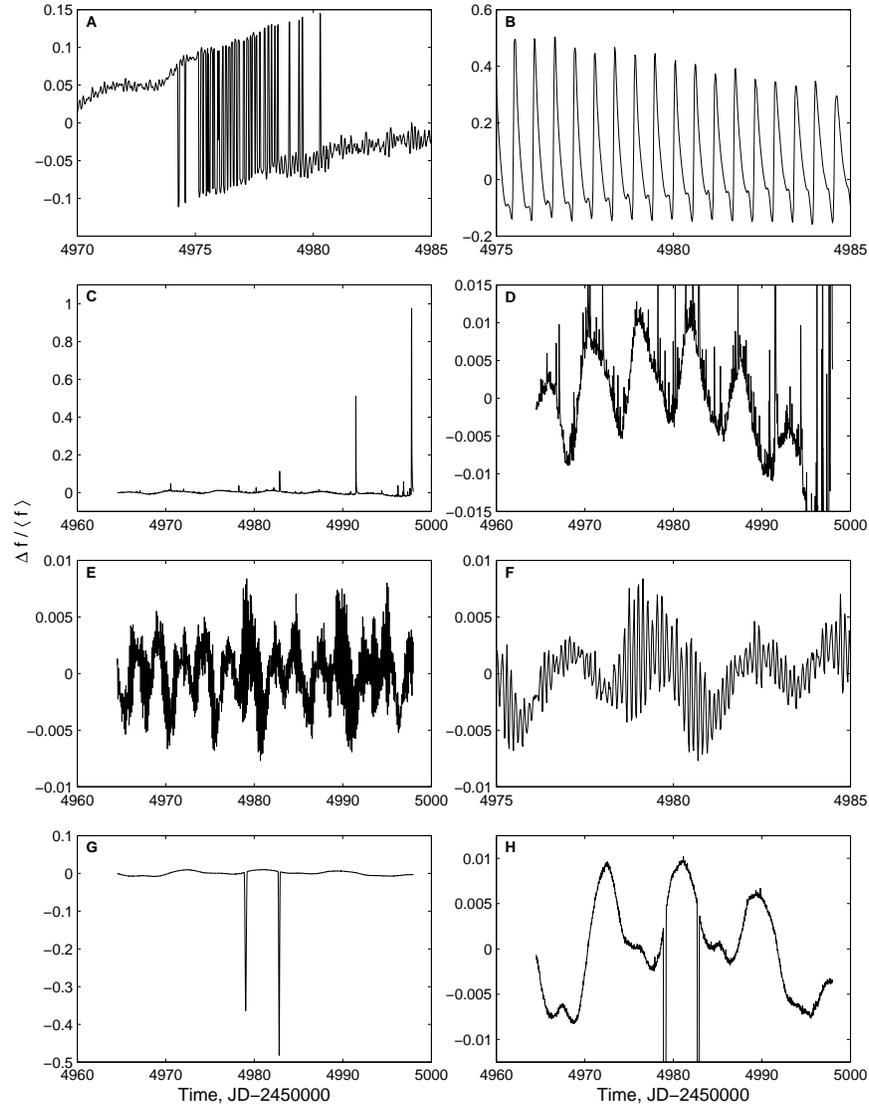}
\caption{Relative flux variations for a variety of large amplitude variable stars and other flux time series exhibiting the broad dynamic range of \emph{Kepler's} sensitivity and stellar variability. The left column displays the entire Q1 light curve, while the right column shows close ups, except where otherwise noted. (A) Light curve for a star whose aperture is located on a column defect and "toggles" as its image drifts.
(B) Close up on an RR-Lyrae star with an evolving envelope. (C and D) A flare star with flares that almost double the flux for this star on top of quasi-periodic variations of a few percent. (E and F) A star exhibiting high frequency oscillations on top of longer term quasi-periodic variations. (G and H) a highly eccentric eclipsing binary.
\label{fig:lightcurves1}}
\end{figure*}

A few otherwise good targets land on column or pixel defects. Secular drift in pointing together with DVA carries stars by up to $\sim$0.7 pixels over each quarter, and as an affected star moves across a column defect, short-term pointing errors can lead to abrupt changes in the measured flux value so that the flux appears to "toggle" between two states for a period of time. Currently we make no attempt to identify and correct this behavior. Panels A and B of Figure \ref{fig:lightcurves1} show the light curve for KID 5286794 illustrating this behavior. Very few targets are affected by this phenomenon, given the paucity of CCD defects, as documented in \citet{caldwell2010}.

\section{Examples of Light Curves}\label{s:lightCurves}
\begin{figure*}
\epsscale{.75}
\plotone{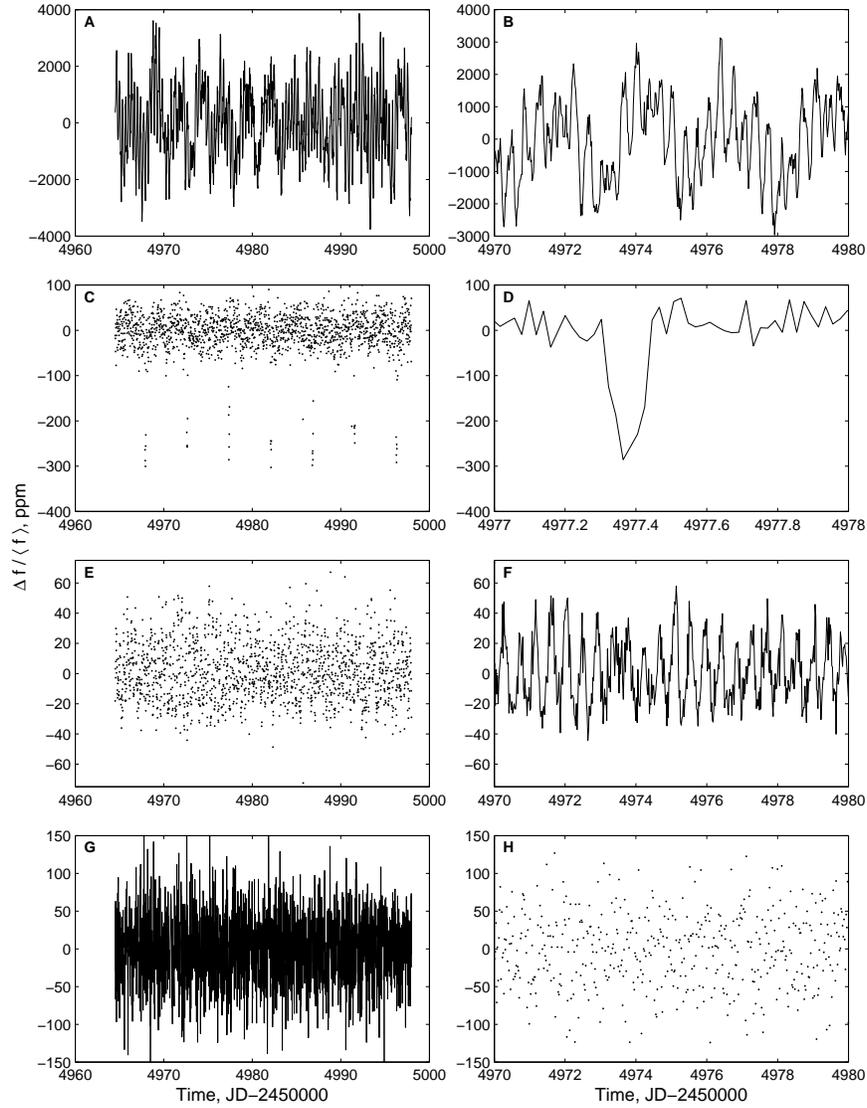}
\caption{Relative flux variations for a variety of small amplitude variable stars and other flux time series exhibiting the broad dynamic range of \emph{Kepler's} sensitivity and stellar variability. The left column displays the entire Q1 light curve, while the right column shows close ups and zooms. (A and B) A star with high frequency oscillations on top of mmag variations. (C and D) Light curve for a star exhibiting 300 ppm transit-like features that can be seen easily by eye. (E and F) A star with variations well below 50 ppm, and coherent, small amplitude oscillations. (G and H) A quiet, main sequence star with a random fluctuations with an rms below 50 ppm on half-hour timescales.
\label{fig:lightcurves2}}
\end{figure*}

In this section we present a potpourri of light curves obtained during Q1 observations that illustrate \emph{Kepler's} stunning photometric precision and the remarkable diversity of stellar photometric behavior across a wide range of amplitudes and timescales. Panels (B) through (H) of Figure \ref{fig:lightcurves1} display large amplitude variable stars' light curves. Low amplitude variable stars are exhibited in Figure \ref{fig:lightcurves2}. The light curve featured in panels (C) and D) is one example of the fact that \emph{Kepler} achieves a precision sufficient to find transit-like features on the order of 200 ppm "by eye".

\section{Conclusions}\label{s:conclusions}
\emph{Kepler's} grand voyage of discovery has begun and in the short while since Science Operations commenced we are already obtaining the photometric precision so indispensable to the task of discovering Earth-like planets transiting solar-like stars. There are a number of instrumental systematics driven by changes in the temperature of the telescope and spacecraft that are both extrinsic to the photometer, such as the seasonal variations due to the heliocentric orbit and quarterly rolls, and intrinsic to hardware on the spacecraft, such as the reaction wheel heater cycles. We are adjusting operational parameters in order to eliminate or reduce the amplitude of intrinsic systematics, or to move the timescale for these instrumental signatures outside of those relevant to transit detection, namely from 2 to 16 hours. These systematic errors represent nuisances that can be filtered out by data processing at the expense of some non-exoplanet astrophysics. The light curves obtained to date by \emph{Kepler} are a delight to behold and promise unforeseen discoveries and a wealth of information about star's photometric behavior at a precision and over a contiguous and extended period of time not heretofore possible. We look forward to establishing a tight constraint on the frequency of earth-like planets, a quantum leap in our understanding of the formation of planetary systems, major strides from asteroseismic results and many serendipitous discoveries  wholly unrelated to exoplanet science.


\acknowledgments
Funding for this Discovery Mission is provided by NASA's Science Mission Directorate. We thank the thousands of people whose efforts made \emph{Kepler} possible.



{\it Facilities:} \facility{Kepler}

\clearpage

\clearpage

\clearpage

\clearpage

\clearpage

\clearpage


\begin{thebibliography}{}

\bibitem[Aigrain et al. (2009)]{aigrain2009} Aigrain, S., et al. 2009, A\&A 506, 425

\bibitem[Batalha et al. (2010)]{batalha2010} Batalha, N. B., et al.   2010, \apj, this issue 


\bibitem[Borucki et al. (2010)] {borucki2010} Borucki, W. J., et al. 2010, Science,
    submitted


\bibitem[Caldwell et al. (2010)]{caldwell2010} Caldwell, D. A., et al.    2010, \apj, this issue


\bibitem[Gilliland (2008)]{gilliland2008} Gilliland, R. L.   2008, \apj, 136, 566

\bibitem[Gilliland et al. (2010a)]{gilliland2010a} Gilliland, R. L.,  et al.  2010a, \apj, this issue

\bibitem[Gilliland et al. (2010b)]{gilliland2010b} Gilliland, R. L.,    et al. 2010b, PASP, in press




\bibitem[Jenkins et al. (2010)]{jenkins2010} Jenkins, J. M., et al. 2010, \apj, this issue

\bibitem[Haas et al. (2010)]{haas2010} Haas, M.,   et al.   2010, \apj, this issue

\bibitem[Koch et al. (2010)]{koch2010} Koch, D. G., et al.     2010, \apj, this issue

\bibitem[Van Cleve \& Caldwell (2009)]{vancleveandcaldwell2009}  Van Cleve, J., \& D. A. Caldwell,     2009, Kepler Instrument Handbook, KSCI 19033-001, (Moffett Field, CA: NASA Ames Research Center)

\end{thebibliography}
\end{document}